\documentclass[prb,twocolumn,superscriptaddress,amssymb,showpacs]{revtex4}
\usepackage{amsmath}
\usepackage{graphicx}

\begin{document}
\title{Complex charge ordering in CeRuSn} 
\author{R. Feyerherm}
\email[Email of corresponding author: ]{ralf.feyerherm@helmholtz-berlin.de} 
\affiliation{Helmholtz-Zentrum Berlin f\"ur Materialien und Energie GmbH, BESSY, 
12489  Berlin, Germany}
\author{E. Dudzik}
\affiliation{Helmholtz-Zentrum Berlin f\"ur Materialien und Energie GmbH, BESSY, 
12489  Berlin, Germany}
\author{S. Valencia}
\affiliation{Helmholtz-Zentrum Berlin f\"ur Materialien und Energie GmbH, BESSY, 
12489  Berlin, Germany}
\author{J. A. Mydosh}
\affiliation{Max Planck Institute for Chemical Physics of Solids, D-01187 Dresden, Germany}
\affiliation{Kamerlingh Onnes Laboratory, Leiden University, 2300RA Leiden, The Netherlands}
\author{Y.-K. Huang}
\affiliation{Van der Waals-Zeeman Institute, University of Amsterdam, 1018XE Amsterdam, The Netherlands}
\author{W. Hermes}
\affiliation{Institute for Inorganic and Analytical Chemistry, University of M\"unster, 48149 M\"unster, Germany}
\author{R. P\"ottgen}
\affiliation{Institute for Inorganic and Analytical Chemistry, University of M\"unster, 48149 M\"unster, Germany}

\date{18.11.2011}

\begin{abstract}
At ambient temperatures, CeRuSn exhibits an extraordinary structure with a coexistence of two types of Ce ions in a metallic environment, namely trivalent Ce$^{3+}$ and intermediate valent Ce$^{(4-\delta)+}$. Charge ordering produces a doubling of the unit cell along the $c$-axis with respect to the basic monoclinic CeCoAl type structure. Below room temperature, a phase transition with very broad hysteresis has been observed in various bulk properties like electrical resistivity, magnetic susceptibility, and specific heat. The present x-ray diffraction results show that at low temperatures the doubling of the CeCoAl type structure is replaced by an ill-defined modulated ground state. In this state, at least three different modulation periods compete, with the dominant mode close to a tripling of the basic cell. The transition is accompanied by a significant contraction of the $c$ axis. XANES data suggest that the average Ce valence remains constant, thus the observed $c$ axis contraction is not due to any valence transition. We propose a qualitative structure model with modified stacking sequences of Ce$^{3+}$ and Ce$^{(4-\delta)+}$ layers in the various modulated phases. Surprisingly, far below 100~K the modulated state is sensitive to x-ray irradiation at photon fluxes available at a synchrotron. With photon fluxes of order $10^{12}$~${\rm s}^{-1}$, the modulated ground state can be destroyed on a timescale of minutes and the doubling of the CeCoAl cell observed at room temperature is recovered. The final state is metastable at 10~K. Heating the sample above 60~K again leads to a recovery of the modulated state. Thus, CeRuSn exhibits both thermally and x-ray induced reversible transformations of the Ce$^{3+}$/Ce$^{(4-\delta)+}$ charge ordering pattern. Such a behavior is unique among any know intermetallic compound.
\end{abstract}

 \maketitle


\section{Introduction} \label{Intro}

Intermetallic compounds based on cerium have long been studied for their novel correlated electron behavior.\cite{Steward, Szytula, SCES} Of particular interest here are the Ce-stannides when combined with a transition metal into a ternary material.\cite{Kim, Schmidt, Gamza} In this large group of compounds exhibiting valence fluctuations, hybridization gapping  and Kondo-lattice antiferromagnetism, equiatomic CeRuSn possesses extraordinary structural, electronic and magnetic features\cite{Riecken:2007, Matar:2007, Mydosh:2011} that are not fully understood.
      
\begin{figure} [t]
\includegraphics* [width=2.8 in] {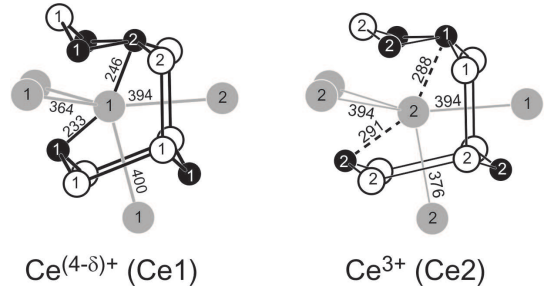}
\caption{Coordination of the two crystallographically independent cerium sites in the room-temperature superstructure of CeRuSn. Cerium, ruthenium, and tin atoms are drawn as medium grey, black filled, and open circles, respectively. The different cerium valences, atom designations, and relevant interatomic distances (pm) are indicated.}
\label{Structure1}
\end{figure}

At ambient temperatures, CeRuSn adopts an unusual crystal structure with two significantly different Ce sites (Figure~\ref{Structure1}). These are arranged in a way to produce a doubling of the unit cell along the $c$-axis with respect to the basic monoclinic CeCoAl type structure. Based on structural considerations, magnetization data, and band structure calculations, the two different Ce sites have been assigned to two different valence states, namely intermediate valent Ce$^{(4-\delta)+}$ (Ce1) and trivalent Ce$^{3+}$ (Ce2).\cite{Riecken:2007, Matar:2007, Mydosh:2011} Notably, the Ce1 sites are characterized by unusually short Ce-Ru bonds which are frequently observed in Ce-Ru based ternary intermetallic compounds.\cite{Hermes:2009,Mishra:2011}  Coexistence of intermediate and trivalent Ce in CeRuSn appears intriguing, since both are embedded in a sea of conduction electrons. Nevertheless, band structure calculations\cite{Matar:2007} confirm that, with the given crystal structure, Ce1 should exhibit intermediate valence while at the Ce2 site $4f$ electron localization results in the Ce$^{3+}$ state. Supporting evidence is the antiferromagnetic transition within one-half of the Ce sites at $T_N = 2.7$~K.\cite{Mydosh:2011} In the course of the present experiments we have checked that the doubling of the CeCoAl type unit cell caused by this charge ordering is stable from room temperature up to at least 800~K, by following the (001) Bragg reflection that is forbidden in the basic cell (not shown).

Below room temperature, a phase transition with very broad hysteresis has been observed in bulk properties like magnetic susceptibility,\cite{Riecken:2007}  specific heat or electrical resistivity.\cite{Mydosh:2011} In the latter measurements the hysteresis extended over the temperature range between 160~K and 290~K, with increased resistivity below the transition. Magnetic susceptibility data have been described with a Curie-Weiss law above 285~K, which indicated that only 50\% of Ce are magnetic. Below the transition the susceptibility is significantly reduced. The transition was tentatively assigned to a crystallographic phase transformation, driven either by a valence transition or by charge-density wave type of ordering.

In the present work we show, mainly based on synchrotron x-ray diffraction data, that at low temperatures the room-temperature doubling of the CeCoAl type structure is replaced by an ill-defined modulated ground state. In this state, at least three different modulation periods compete, with the dominant mode close to a tripling of the basic CeCoAl cell. The transition is accompanied by a significant contraction of the $c$ axis, which may suggest an increase of the fraction of Ce$^{(4-\delta)+}$ on cooling. However, XANES data suggest that the average Ce valence remains constant. We propose a qualitative picture for the various modulated states, which implies modified stacking sequences of Ce1 and Ce2 layers in the various modulated phases.

In addition to clarifying the low-temperature structural modulation, we observed a surprising x-ray induced effect. Far below 100~K the modulated state is sensitive to x-ray irradiation at photon fluxes available at a synchrotron. With photon fluxes of order $10^{12}$~${\rm s}^{-1}$, the modulated ground state can be destroyed on a timescale of minutes. The doubling of the CeCoAl cell observed at room temperature is recovered. The final state is metastable at 10~K. Heating the sample above 60~K again leads to a recovery of the modulated state.


\section{Experimental}

Two different CeRuSn crystals were investigated in the present work. Preliminary synchrotron x-ray diffraction measurement were carried out on a small CeRuSn single crystal (sample~\#1) of dimensions $\approx 0.02 \times 0.1 \times 0.1$~mm$^3$ that was picked from a polycrystalline batch produced at the University of M\"unster by techniques described earlier.\cite{Riecken:2007} A second, large single crystal (sample~\#2) has been grown by means of the tri-arc Czochralski technique in a Ti-gettered ultra-pure Ar atmosphere at Amsterdam. 
From this large crystal a piece of dimensions $\approx 1 \times 1.5 \times 3$~mm$^3$ was cut and attached to a copper sample holder. After the first thermal cycling between 300~K and 10~K this sample cleaved and developed a shiny face normal to the $c$ axis. Essentially, the x-ray diffraction results on both crystals are consistent, with some quantitative differences, especially of the relative intensities of the various types of low-$T$ superstructure reflections. Where not mentioned otherwise, the data presented in this work were collected on sample~\#2 after cleavage. A small piece of this crystal was used for laboratory x-ray diffraction. After several thermal cycles the rocking width of crystal~\#2 was still  $< 0.5 ^\circ$, indicating good sample quality.

X-ray diffraction has been carried out first on a conventional laboratory four-circle x-ray diffractometer equipped with a displex cryostat using Mo K$_\alpha$ radiation ($\lambda = 0.71073$~\AA) and a scintillation counter. In this stage of the experiment the lattice parameters at 10~K and 325~K were determined by carefully centering more than 50 reflections. We obtained $a=11.552(15)$~\AA, $b=4.748(3)$~\AA, $c=10.146(7)$~\AA, $\beta = 103.49(8)^\circ$ at 10~K and $a=11.565(18)$~\AA, $b=4.759(3)$~\AA, $c=10.222(9)$~\AA, $\beta = 103.19(10)^\circ$ at 325 K. The latter values are in good agreement with literature data for 300~K,\cite{Riecken:2007} where the space group $C2/m$ with $Z = 8$ has been determined. A comparison of the 325~K and 10~K values results in a volume contraction $\Delta V/V = 1.10\%$ on cooling which apparently is dominated by the reduction of the $c$-axis parameter, $\Delta c^*/c^* = 0.87\%$ (where $c^* = c \sin\beta$).

All other x-ray diffraction measurements have been carried out at the synchrotron beamline MAGS\cite{MAGS} at the HZB synchrotron source BESSY using various photon energies between 5.7~keV ($\lambda = 2.175$~\AA) 
and 12.398~keV ($\lambda = 1$~\AA). Again, a displex cryostat and a scintillation counter were employed. These experiments were complemented by x-ray absorption near-edge structure (XANES) measurements at the Ce L$_{3}$ edge (5.72~keV) carried out in fluorescence yield mode at the same beamline. For the latter measurements the photon energy was calibrated with a Cr foil at the Cr K$_\alpha$ edge (5.989 keV).\\

\section{Results and discussion}

\subsection{Structural transition}

\begin{figure} [b]
\includegraphics* [width=3.2 in] {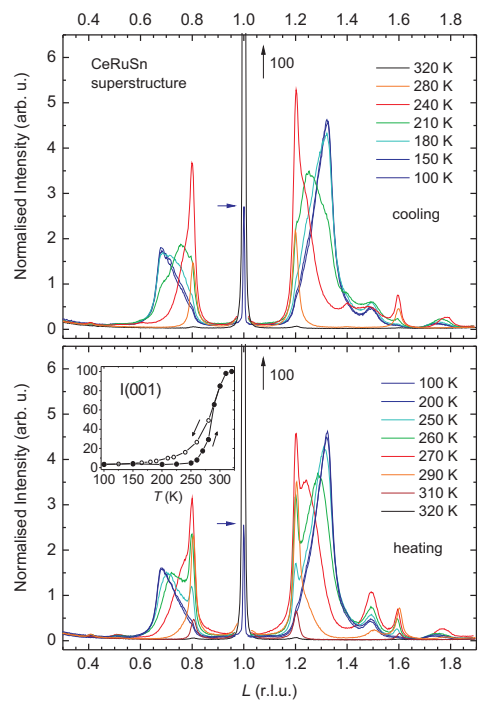}
\caption{Longitudinal scans along the $(00L)$ reciprocal space direction at various temperatures between 100~K and 320~K on cooling (top) and heating (bottom) show the development of various types of superstructure reflections. Between these two data sets, the sample was cooled to 10~K without x-ray irradiation of the sample. The central $(001)$ reflection is strongly suppressed at low temperatures while sets of superstructure reflections with dominant peaks at $L = 0.8, 0.75, 0.68$ and symmetric positions  $L = 1.2, 1.25, 1.32$ appear (and partly disappear again) on cooling. The width of the $(001)$ reflection (marked by arrows) marks the experimental resolution, thus the other superstructure reflections are strongly broadened. The integrated intensity of the low-temperature spectrum is about the same as the intensity of the $(001)$ reflection at 320~K. The $L$ scale is normalized to the $(001)$ position for the given temperatures. The inset shows the temperature dependence of the (001) integrated intensity on cooling and heating.}
\label{Waterfall}
\end{figure}

Preliminary x-ray diffraction data suggested a structural modulation along the $c$ axis for temperatures far below 300~K. To determine the temperature evolution of the corresponding propagation vector we measured longitudinal scans along the $(00L)$ reciprocal space direction for a full temperature cycle 320~K $\rightarrow$ 100~K $\rightarrow$ 320~K (see Figure~\ref{Waterfall}) and found a quite unique behavior.

On cooling, first a set of relatively sharp superstructure reflections at $\tfrac{1}{5}$-integer positions like $L = 0.8, 1.2, 1.6$ develops immediately below ambient temperatures, pointing to a commensurate modulation of the crystal structure that involves a quintupling of the basic CeCoAl type cell. This is followed by a shift of the spectral weight to a set of very broad peaks at $\tfrac{1}{4}$-integer positions $L = 0.75, 1.25, 1.5, 1.75$ which is dominant at 210~K. Finally, these two sets are partially suppressed and a third, incommensurate set at positions $L = 0.68, 1.32$ becomes dominant below 180~K. Note that the latter positions are close to $\tfrac{1}{3}$-integer values. The ground state is ill-defined, since apparently at least the three different modulation wave vectors mentioned above coexist at 100~K. The resulting diffractogram is quite fuzzy, pointing to short coherence lengths or a broad distribution of propagation vectors between 0.68 and 0.8~$c^*$.

On heating, the incommensurate modulation persists up to about 250~K, where the $\tfrac{1}{5}$-integer set of reflections starts to develop again. At 270~K we observe a coexistence of $\tfrac{1}{5}$- and $\tfrac{1}{4}$-integer reflections, while above 290~K up to 320~K only $\tfrac{1}{5}$-integer peaks are found. Comparing this sequence with the cooling data, we find a very broad hysteresis range between 150~K and 300~K. This range is roughly the same as that observed in the measurements of the bulk properties of a polycrystalline sample.\cite{Mydosh:2011} 

Notably, the intensity of the central $(001)$ reflection is strongly diminished at low temperatures. Actually, at 100~K it is only about 3\% of the intensity at 320~K. The inset in Figure~\ref{Waterfall} shows the temperature dependence of the (001) intensity on cooling and heating. The latter curve clearly marks the 290~K transition observed in the bulk properties. This transition corresponds to the crossover from the relatively well-defined $\tfrac{1}{5}$- integer modulation to the high-temperature phase, which, however, is completed only at 320~K. Interestingly, the integrated intensity of the low-temperature spectrum is roughly (to within 10\%) the same as the intensity of the $(001)$ reflection at 320~K. Thus, at low $T$ the $(001)$ reflection appears to be replaced by the observed set of new superstructure reflections.


\begin{figure} 
\includegraphics* [width=3.2 in] {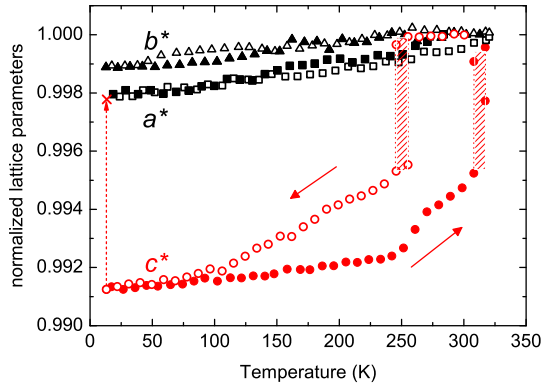}
\caption{(Color online) Temperature dependence of $d$ spacings for the three principal reciprocal space directions normalized to their value at 320~K for sample~\#1. The data points were measured with strong attenuation of the primary beam. Thus, any irradiation induced effects below 100~K, as discussed in Sect.~\ref{subsect}, should be absent.
Shaded areas mark regions of coexistence of two phases. The dashed arrow shows the x-ray irradiation induced expansion of $c^*$ as discussed in Sect.~\ref{subsect}.}
\label{LatticeParameters}
\end{figure}

It is interesting to compare these results with the temperature evolution of the lattice parameters, which were determined from the Bragg angles for the three principal reflections (400), (020), and (004) on sample~\#1 between 10~K and 320~K during heating and cooling. Figure~\ref{LatticeParameters} presents the corresponding $d$ spacings normalized to their value at 320~K. While $a^*$ and $b^*$ show only a small and steady temperature dependence, $c^*$ exhibits large jumps at about 250~K on cooling and at about 310~K on heating. Here, hysteresis effects appear to extend over the whole temperature range between 100~K and 310~K. We observe at least 10~K broad temperature regions of coexistence of two phases in both cases (see shaded areas in Figure~\ref{LatticeParameters}). 

\begin{figure} 
\includegraphics* [width=3.2 in] {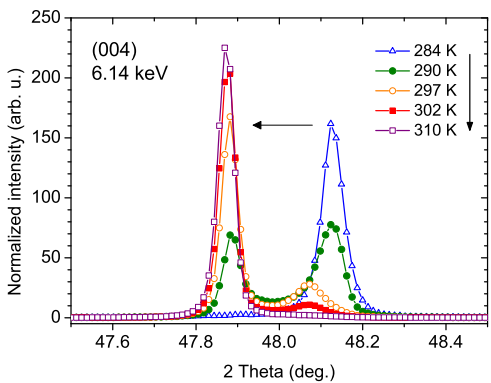}
\caption{(Color online) Longitudinal scans of the (004) Bragg reflection in the transition region. Splitting of this reflection indicates coexistence of two phases. A comparison with Figure~\ref{Waterfall} suggests that the component at lower angle corresponds to the high-temperature phase and the component at higher angle to the phase associated with the $\tfrac{1}{5}$-integer set of superstructure reflections. Peak intensities are area normalized.}
\label{Splitting}
\end{figure}

The phase coexistence is exemplified in Figure~\ref{Splitting}, which shows the splitting of the (004) Bragg reflection in the transition region for sample~\#2. Comparison with Figure~\ref{Waterfall} suggests that the two components correspond to the high-temperature phase and the phase associated with the $\tfrac{1}{5}$-integer set of superstructure reflections. From Figures~\ref{LatticeParameters} and \ref{Splitting} we infer a jump by 0.45\% of $c^*$ directly at the transition between these two phases, where the $\tfrac{1}{5}$-phase has the smaller $c^*$ parameter. Correspondingly, the kink at 250~K in the heating curve (cf. Figure~\ref{LatticeParameters}) can be associated with the onset of the transition from the ground state to the $\tfrac{1}{5}$-phase. We infer that the $c^*$ parameter of the phase associated with the $\tfrac{1}{3}$-integer set of superstructure reflections is about 0.25\% smaller than that of the $\tfrac{1}{5}$-phase.  We note that the total difference of the $c^*$ parameter of these two phases with respect to the high-temperature phase, namely 0.45\% and 0.70\%, scale well with the difference $\delta Q$ of the corresponding propagation vectors, $1-0.8 = 0.2$ and $1 - 0.68 = 0.32$, respectively. This indicates that the modulated phases are characterized by regular arrangements of structural units with collapsed extension along the $c$ direction, where the number of collapsed units scales with $\delta Q$. We will return to this point below.

\begin{figure} [b]
\includegraphics*  [width=3.2 in]  {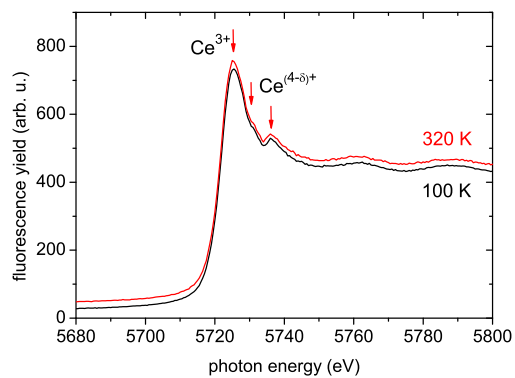}
\caption{(Color online) Fluorescence spectra at the Ce L$_3$ edge. The ratio of intensities of the spectral features indicated by arrows allows an estimate of the Ce$^{3+}$/Ce$^{4+}$ ratio and thus of the average Ce valence. Any increase of the average Ce valence would lead to a significant reduction of the peak at 5725~eV, in contrast to the observation. The two curves measured at 100~K and 320~K, respectively, are offset for clarity.}
\label{Spectroscopy}
\end{figure}

Since in CeRuSn the Ce1~=~Ce$^{(4-\delta)+}$ sites are characterized by shortened Ce-Ru bonds which mainly extend along the $c$ direction,\cite{Riecken:2007} the observed $c$ axis contraction may suggest an increase of the fraction of Ce$^{(4-\delta)+}$ on cooling. This would imply a change of the average Ce valence. To determine the Ce valence in CeRuSn, we carried out XANES measurements. Figure~\ref{Spectroscopy} shows the fluorescence spectra measured around the Ce L$_3$ edge at 100~K and 320~K, i.e., well below and above the transition respectively. Three spectroscopic features are observed close to the edge, that are well known from intermediate-valence or mixed-valence Ce compounds. The lowest energy feature results from Ce$^{3+}$ with the $4f^1$ final state while a pair of peaks at higher energies is related to Ce$^{(4-\delta)+}$ with $4f^1\underline{L}$ and $4f^0$ final states.\cite{Bianconi} The relative intensities of these features can be determined by standard procedures. We obtain an average Ce valence of 3.18, corresponding to a valence 3.36 for Ce1. Note that, e.g., in CeO$_2$ the formal Ce$^{4+}$ state has a valence of about 3.5 only.\cite{Niewaa} No significant temperature variation is observed between 100~K and 320~K in CeRuSn, pointing to a temperature independent average Ce valence. Thus, the observed $c$ axis contraction of the modulated phases appears {\textit not} to be associated with any valence transition.

\begin{figure} [t]
\includegraphics* [width=2.5 in] {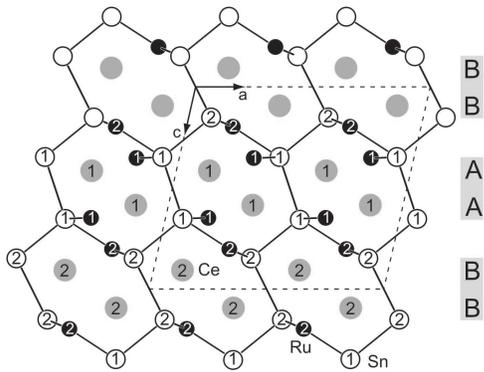}
\caption{Projection of the commensurate superstructure of CeRuSn (room temperature phase) onto the $xz$ plane. Cerium, ruthenium, and tin atoms are drawn as medium grey, black filled, and open circles, respectively. Atom designations for the crystallographically independent sites are given at the bottom of the drawing. The AA and BB packages of Ce$^{(4-\delta)+}$ and Ce$^{3+}$ layers are emphasized at the right-hand part. For details see text.}
\label{Structure2}
\end{figure}

The present experimental results suggest a quite unique behavior of CeRuSn. Even the reported room temperature structure, space group $C2/m$, is unconventional, since it results from a doubling of the CeCoAl type basic structure along the $c$ axis induced by charge ordering of Ce$^{(4-\delta)+}$ (Ce1) and Ce$^{3+}$ (Ce2) ions in form of a stacking sequence (..$|$AABB$|$AABB$|$..) along $c$, where "A" and "B" stand for a pair of nearest-neighbor Ce1 and Ce2, respectively (Figure~\ref{Structure2}). In this sense, the $(001)$ reflection may already be regarded as superstructure reflection resulting from charge ordering and the associated modifications of the individual Ce environment (such as short Ce-Ru bonds).

On cooling, close to room temperature, apparently a first order type of crystallographic phase transition takes place in which the doubling of the CeCoAl type cell breaks down and is replaced by other, more complex modulations along the $c$ axis which, however, may be closely related to the original superstructure. We suppose that the superstructures observed at low $T$ still involve a distribution of the two types of Ce sites within the lattice. 
Since the ground state is ill-defined, detailed structural information for the various low temperature phases can not be obtained. However, we arrive at a qualitative picture by a more careful look at the various sets of superstructure reflections. First, we note that the positions of the incommensurate reflections, 0.68 and 1.32, are close to the rational values $\tfrac{2}{3}$ and $\tfrac{4}{3}$. These positions would be compatible with a propagation vector $q = \tfrac{2}{3}~c^*$, i.e., a tripling of the CeCoAl type basic structure. As the simplest model consistent with the observed modulation we propose a stacking sequence (..$|$AAABBB$|$AAABBB$|$..), with "A" and "B" defined above. Assuming that pairs "AA" of neighboring Ce1 layers have a reduced spacing along $c$ compared to "AB" or "BB" pairs, this structure model would explain the observed contraction of $c^*$ with respect to the high-temperature phase. In the latter, the fraction of "AA" pairs is $1/4$, while in the proposed structure for $q = \tfrac{2}{3}~c^*$ it increases to $1/3$.

For the $\tfrac{1}{5}$-integer set we note that the dominant reflections at $L = 0.8, 1.2$ are accompanied by smaller reflections at $L = 0.4$ and $1.6$. 
This observation suggests a modulation wave vector $q = \tfrac{2}{5}~c^*$, i.e. a quintupling of the basic structure, but is incompatible with a simple sinusoidal modulation, in which the (higher harmonic) reflections at $L = 0.8, 1.2$ would be absent. Instead, calculating the Fourier transforms of several possible stacking models we found that the observed set of $\tfrac{1}{5}$-integer reflections may be interpreted as resulting from a stacking sequence (..$|$AAABBBAABB$|$AAABBBAABB$|$..). Here, the fraction of of "AA" pairs is $3/10$.  We also note that for the intermediate $\tfrac{1}{4}$-integer set of reflections at $L =$ 0.5, 0.75 (1.25, 1.5, 1.75) the intensity  increases (decreases) in that order. This observation again is incompatible with a sinusoidal modulation but is well reproduced assuming a stacking sequence (..$|$AAABBBAAABBBAABB$|$..) with propagation vector $q = \tfrac{1}{4}~c^*$ and fraction of "AA" pairs $5/16 = 0.3125$. 

The proposed stacking sequences follow a simple scheme, namely the successive introduction of extra "AA" and "BB" pairs on cooling, see Table~\ref{Table1}. The increase of the fraction of "AA" (or "BB") pairs $0.25 \rightarrow 0.3 \rightarrow 0.3125 \rightarrow 0.333$ scales perfectly with the positions of the dominant superstructure reflections for each phase, $1 \rightarrow 1.2 \rightarrow 1.25 \rightarrow 1.333$. This simple match gives us confidence that our qualitative model is reasonable.

\begin{table} [t]
\caption{Suggested stacking sequences associated with the observed superstructure reflections. Values of $q$ are related to the room $T$ lattice of CeRuSn. The $q = \tfrac{2}{3}$ stacking is idealized, the observed value is $q = 0.68$.}
\label{Table1}
\begin{tabular}{llc}
\hline\noalign{\smallskip}
strongest & proposed stacking & $q/c^*$\\
reflections ($L$) & sequence & \\
\noalign{\smallskip}\hline\noalign{\smallskip}
1    & $|$AABB$|$AABB$|$ & 0\\
0.8, 1.2  & $|$AAABBBAABB$|$AAABBBAABB$|$ & $\tfrac{2}{5}$ \\
0.75, 1.25 & $|$AAABBBAAABBBAABB$|$ &$\tfrac{1}{4}$\\
0.667, 1.333 & $|$AAABBB$|$AAABBB$|$ &$\tfrac{2}{3}$\\
\noalign{\smallskip}\hline
\end{tabular}
\end{table}

\subsection{X-ray radiation induced effects} \label{subsect}

While exploring the temperature region far below 100~K without attenuation of the primary synchrotron x-ray beam, we noticed a surprising effect: On a time scale of minutes the modulated ground state is destroyed while the commensurate room temperature state appears to be recovered. This prevented us from measuring the modulated state at temperatures well below 100~K.

\begin{figure} 
\includegraphics* [width=3.2 in] {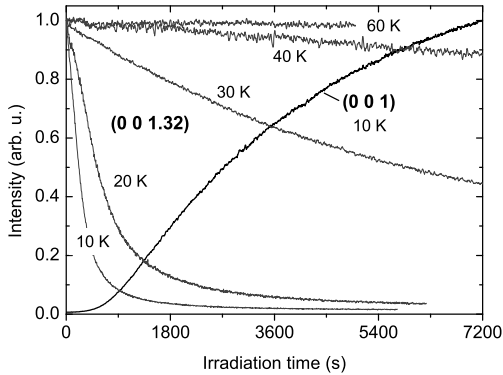}
\caption{Intensity at (001) and (0~0~1.32) reciprocal space positions as function of irradiation time with $E = 5724$~eV photons at full available flux ($\approx 5 \times $ 10$^{11}$ s$^{-1}$, taking into account absorption through the Be domes of the cryostat) measured at 10~K. The sample was cooled to 10~K from 300~K with closed photon shutter in order to start from the fully developed modulated state. The beam footprint on the sample was $\approx 1$~mm$^{-2}$, the attenuation length is $\approx 3 \mu$m for 5724~eV photons in CeRuSn. 
The decay time for the (0~0~1.32) intensity is temperature dependent. The time constants $t_{1/2}$ are 220~s, 540~s, 1.7~h, 11~h, and $>280$~h at 10~K, 20~K, 30~K, 40~K, and 60 ~K, respectively, resulting in a behavior $\ln (t_{1/2}) \propto T$. }
\label{Burn}
\end{figure}

Figure~\ref{Burn} shows the intensity at (001) and (0~0~1.32) reciprocal space positions as function of irradiation time at full available flux for photon energy $E = 5724$~eV (i.e., at the Ce L$_3$ edge) measured at 10~K. After a short initial period of somewhat slower decay, the intensity of the (0~0~1.32) reflection, characteristic for the modulated ground state, shows a steep decrease roughly linear with irradiation time. The intensity is reduced to 1/2 of the initial value within $t_{1/2} = 220$~s. The time dependence is similar to an exponential decay but can not be described by a simple exponential. Simultaneously with the decay of the (0~0~1.32) intensity, the (001) starts to grow. After about 2~h its intensity saturates at  value similar to the intensity observed at room temperature. Starting from the resulting state, we have followed the (001) intensity at 10~K with the x-ray intensity reduced by a factor of 25 using an appropriate attenuator. Within four hours the (001) intensity was reduced by only 35\% (not shown). Assuming a roughly exponential decay, this corresponds to a time constant $t_{1/2}$ of about 6.4~hours. This shows that the final state after irradiation is metastable at 10~K.

\begin{figure} 
\includegraphics* [width=3.2 in] {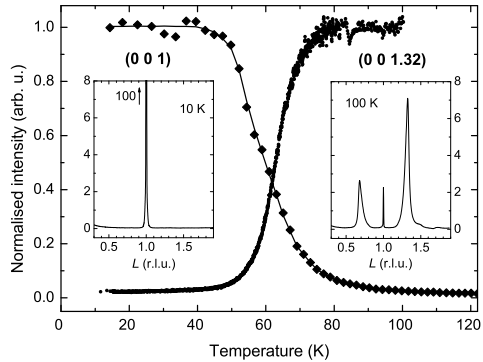}
\caption{Integrated intensity of the (001) reflection from $L$-scans and intensity at the (0~0~1.32) reciprocal space position as function of temperature on slow heating ($\approx 2$~K/min) after 2~h of irradiation with full available flux for photon energy $E = 5724$~eV at 10~K. A transition is observed at 60~K, which leads to a decay of the irradiation-induced state and a recovery of the modulated ground state. During heating, the sample was still fully irradiated. Insets show longitudinal scans along the $(00L)$ reciprocal space direction; (left) after 2~h of full irradiation at 10~K, where only the (001) reflection is present, suggesting that the modulated ground state is destroyed and (right)  after reaching 100~K, where the modulated state is recovered.}
\label{Annealing}
\end{figure}

We have repeated the x-ray irradiation procedure at increasing temperatures and followed the (0~0~1.32) intensity, see Figure~\ref{Burn}. Interestingly, already at 30~K the modulated ground state is much less affected by irradiation than at 10~K and above 60~K the modulated state appears to be completely stable against irradiation. The time constants roughly follow the equation $\ln (t_{1/2}) \propto T$. This shows that the observed effect is not due to local heating of the sample by the x-ray beam, which was not expected in any case since CeRuSn is metallic (with associated high thermal conductivity) and the heat input by the beam ($\approx  0.5$~mW) is much smaller than the cooling-power of the cryostat. Rather, irradiation appears to create a metastable state that at very low temperatures is frozen in but at elevated temperatures relaxes back to the ground state by thermal activation. Consequently, one would expect that the irradiation-induced state can be "annealed" by heating. This is indeed observed in the experiment: Figure~\ref{Annealing} shows the temperature dependence of the intensity of the (001) and (0~0~1.32) reflections on slowly heating the sample, starting from the fully irradiated state. We observe a transition at 60~K above which the (001) breaks down while simultaneously the (0~0~1.32) reflection is recovered. Figure~\ref{Annealing} (insets) show longitudinal scans along the $(00L)$ reciprocal space direction after 2~h of full irradiation measured at 10~K and 100~K.  At 10~K, only the (001) reflection is present, suggesting that the modulated ground state is completely destroyed while after reaching 100~K, the modulated state is recovered. 

\begin{figure} [t]
\includegraphics* [width=3.0 in] {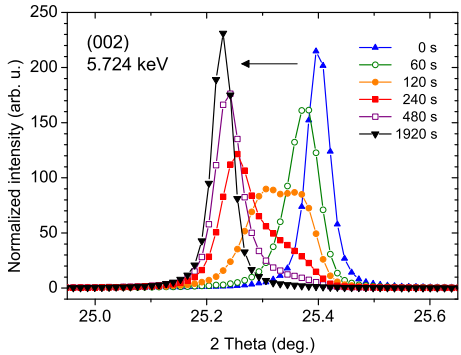}
\caption{(Color online) Evolution of the (002) reflection on irradiation at 10~K for increasing total irradiation times at full available photon flux. Already after one minute of irradiation the position and shape of the reflection changes. After a few minutes the reflection splits. At the end of the irradiation period only a single peak at lower Bragg angle $2 \theta$, corresponding to an increased $c^*$ parameter, remains. The total shift corresponds to an increase of $c^*$ by 0.64\%. Comparison with Figure~\ref{LatticeParameters} shows that this increase completely compensates the effect of the phase transition to the modulated ground state. While the irradiation was carried out with full flux, the data were measured with beam intensity reduced to $3 \times 10^{-3}$.}
\label{002}
\end{figure}

Remarkably, the re-appearance of the (001) reflection on irradiation is accompanied by an increase of the $c^*$ parameter by 0.64\%. This is demonstrated by Figure~\ref{002} which shows the evolution of the (002) reflection on irradiation. Interestingly, the observed 0.64\% increase on irradiation brings the $c^*$ parameter to a value which is quite in accordance with the general thermal expansion effect observed for $a^*$ and $b^*$ (cf. Figure~\ref{LatticeParameters}). This supports our conclusion that massive x-ray irradiation at 10~K appears to destroy the modulated ground state while restoring the room temperature state.

Finally we have checked the irradiation procedure at different photon energies and attenuator settings (not shown). At photon energy $E = 5700$~eV, slightly below the Ce L$_3$ absorption edge (attenuation length $\mu = 2.9 \mu$m),\cite{Atten} the decrease of the (0~0~1.32) intensity is about a factor of 2.5 slower ($t_{1/2} = 570$~s) than at  $E = 5724$~eV ($\mu = 2.1 \mu$m). At photon energy $E = 12.4$~keV ($\mu = 17 \mu$m),  even with a four times higher photon flux of $\phi \approx 2 \times $ 10$^{12}$ s$^{-1}$, the effect is still slower ($t_{1/2} = 710$~s). Attenuating the beam intensity by a factor of six 
at $E = 12.4$~keV, we found a six times slower decay than without attenuation. This behavior can be understood considering that the absorbed energy per volume is largest at the absorption edges and smallest at high photon energies. Any irradiation induced effect should scale with the absorbed energy per volume. This consideration also explains the non-exponential decay of the (0~0~1.32) intensity on irradiation. Due to absorption, lower-lying parts of the sample are irradiated with less beam intensity and thus will show a slower decay. The time dependence of the intensity would therefore be best described by a distribution of exponential decay times (a detailed quantitative analysis of the time dependence is beyond the scope of this work). We note that quantitatively, $t_{1/2}$ scales with $\mu / \phi$ when comparing the effect at 5700 eV and 12.4 keV, but directly at the Ce L$_3$ absorption edge the (0~0~1.32) intensity decays about a factor of 2 faster than expected. Thus, enhanced absorption directly at the Ce sites speeds up the irradiation-induced phase transformation.


\section{Summary and Conclusions}

We have investigated the structure and valence properties of single-crystal CeRuSn using synchrotron x-ray diffraction and XANES. This intermetallic compound becomes structurally modulated through a series of first-order stacking transitions along its $c$-axis. Beginning from at least 800~K down to ambient temperature the unit cell of the $C2/m$ space group is doubled along $c$ with respect to the CeCoAl reference structure. This produces two inequivalent sites for Ce, Ru, and Sn, thereby generating charge ordering of two Ce valences, i.e., Ce$^{3+}$ and Ce$^{(4-\delta)+}$. As the temperature is reduced below ambient, further modulations occur, viz., quintupling, quadrupling and finally an incommensurate modulation close to a tripling below 180~K.

The temperature dependence of the XANES spectra at the Ce-L$_3$ edge show that the Ce-valence ratio Ce$^{3+}$/Ce$^{(4-\delta)+}$ remains constant throughout the transitions. Accordingly we propose qualitative model structures for the stacking sequences in the various modulated phases based upon equal fractions of Ce$^{3+}$ and Ce$^{(4-\delta)+}$. Irradiating the sample with high photon flux at 10~K, the stacking sequence reverts back to the ambient temperature doubled modulation. However, at 60~K and above, thermal activation rapidly regenerates the tripled ground state. We have attempted to interpret the structural behavior in terms of digitized stacking faults locking-in the modulation with integer sequences. The unusually low energy driving the modulations ($<25$~meV) in CeRuSn is likely caused by magneto-elastic coupling due to the mixed magnetic moments and valence fluctuations.

The present x-ray diffraction result show that CeRuSn exhibits both thermally and x-ray induced transformations of the Ce$^{3+}$/Ce$^{(4-\delta)+}$ charge ordering pattern. To our knowledge, such a complex behavior is unique among any know intermetallic compound. Related \textit{reversible} x-ray induced effects have been observed before only in some manganites~\cite{Kiryukhin} and in CuIr$_2$S$_4$.\cite{Ishibashi:2002} In the manganites, charge-ordering results in an antiferromagnetic insulating ground state. X-ray illumination at low temperatures leads to a destruction of the charge-ordered state resulting in the recovery of the room-temperature metallic ferromagnetic state. This effect can be reversed by thermal cycling, i.e., the insulating ground state is recovered above about 60~K (by chance the same temperature as observed in CeRuSn, see Fig.~\ref{Annealing}). In 
CuIr$_2$S$_4$, the insulating ground state is characterized by the formation of Ir$^{4+}$ dimers which are destroyed by x-ray illumination. Again, the final state is conducting and thermal cycling to above 70~K recovers the insulating ground state. In both types of compounds, a strong electron-lattice interaction was held responsible for the observed behavior. X-ray irradiation leads to a change of the local electronic structure and consequently the lattice relaxes, producing a metastable state. A similar mechanism appears to act in CeRuSn.  However, CeRuSn is unique in exhibiting transformations between various different charge ordered states, while the other compounds only order in a single pattern and become disordered on heating or irradiation.

To conclude, CeRuSn exhibits a unique behavior and thus adds a new flavor to the already very rich spectrum of electronic ground states observed in cerium based correlated electron systems.\\




\textit{We thank U. Burkhardt, H. Borrmann (MPI-CPfS, Dresden), R.-D. Hoffmann, and U. Ch. Rodewald (IAAC, Universität Münster) for help with the crystal orientation and shaping. JAM wishes to acknowledge useful discussions with U. R\"o{\ss}ler (IFW-Dresden)}

\end{document}